\documentclass[sigconf]{cidr-2023}
\AtBeginDocument{%
  \providecommand\BibTeX{{%
    \normalfont B\kern-0.5em{\scshape i\kern-0.25em b}\kern-0.8em\TeX}}}

\usepackage[ruled, linesnumbered]{algorithm2e}
\SetKwInOut{Input}{Input}
\SetKwInOut{Output}{Output}
\SetKwComment{Comment}{/* }{ */}
\usepackage{enumitem}
\usepackage{graphicx}
\usepackage{xspace}

\newif\ifnotes
\notestrue

\definecolor{tomato}{rgb}{1,0.2,0}
\definecolor{grey}{rgb}{0.4,0.4,0.4}

\newcommand{\sysname}{\mbox{\sc WarpGate}\xspace}
\newcommand{\plainsysname}{WarpGate\xspace}
\newcommand{\nextiaJD}{Nextia\textsubscript{JD}\xspace}

\newcommand{\code}[1]{\texttt{#1}}
\newcommand{\ignore}[1]{}
\newcommand{\subhead}[1]{\vspace{0.3\baselineskip}\noindent\textbf{#1}}

\theoremstyle{definition}
\newtheorem*{definition}{Definition}
\newtheorem*{ps}{Problem Statement}

\begin{document}

\title{\plainsysname: A Semantic Join Discovery System \\for Cloud Data Warehouses}

\author{Tianji Cong$^*$}
\affiliation{
  \institution{University of Michigan}
  \city{Ann Arbor}
  \state{Michigan}
  \country{USA}
}
\email{congtj@umich.edu}
\thanks{$^*$Work done when Tianji was interning at Sigma Computing}

\author{James Gale}
\affiliation{
  \institution{Sigma Computing}
  \city{San Francisco}
  \state{California}
  \country{USA}
}
\email{jlg@sigmacomputing.com}

\author{Jason Frantz}
\affiliation{
  \institution{Sigma Computing}
  \city{San Francisco}
  \state{California}
  \country{USA}
}
\email{jason@sigmacomputing.com}

\author{H. V. Jagadish}
\affiliation{
  \institution{University of Michigan}
  \city{Ann Arbor}
  \state{Michigan}
  \country{USA}
}
\email{jag@umich.edu}

\author{\c{C}a\u{g}atay Demiralp}
\affiliation{
  \institution{Sigma Computing}
  \city{San Francisco}
  \state{California}
  \country{USA}
}
\email{cagatay@sigmacomputing.com}


\begin{abstract}
  Data discovery is a major challenge in enterprise data analysis: users often struggle to find data relevant to their analysis goals or even to navigate through data across data sources, each of which may easily contain thousands of tables. One common user need is to discover tables joinable with a given table. This need is particularly critical because join is a ubiquitous operation in data analysis, and join paths are mostly obscure to users, especially across databases. Furthermore, users are typically interested in finding ``semantically'' joinable tables: with columns that can be transformed to become joinable even if they are not joinable as currently represented in the data store. 

  We present \sysname, a system prototype for data discovery over cloud data warehouses. \sysname implements an embedding-based solution to semantic join discovery, which encodes columns into high-dimensional vector space such that joinable columns map to points that are near each other. Through experiments on several table corpora, we show that \sysname (i) captures semantic relationships between tables, especially those across databases, and (ii) is sample efficient and thus scalable to very large tables of millions of rows. We also showcase an application of \sysname within an enterprise product for cloud data analytics.
\end{abstract}

\maketitle

\begin{figure*}[h!t]
  \centering
  \includegraphics[width=0.95\textwidth]{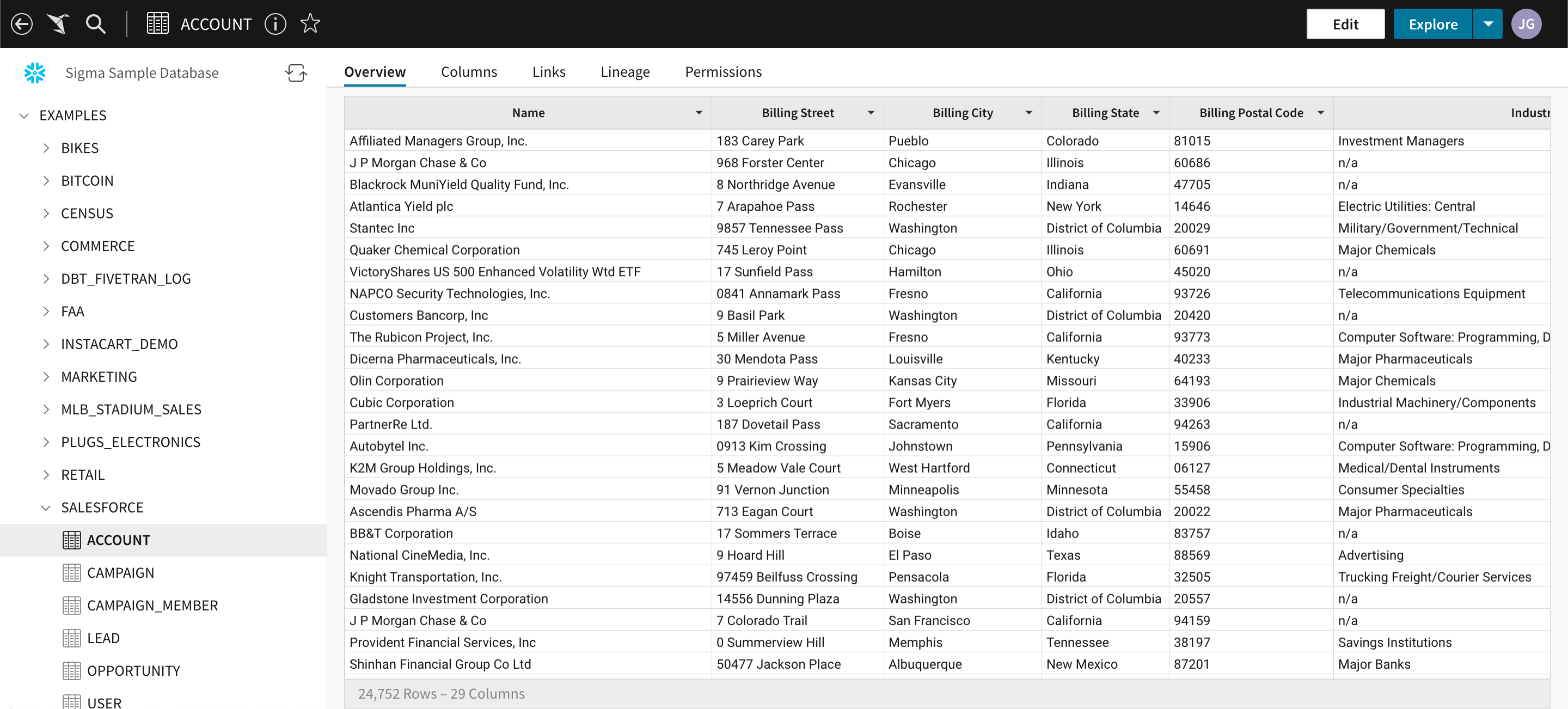}
  \caption{The interface of Sigma Workbooks connecting to a CDW. Consider a business user Joey who is interested in launching a sales campaign and identifying ideal customers. She first opens up the \code{ACCOUNT} table in their \code{SALESFORCE} database, which lists all customer companies. Next, she wants to augment the table with relevant columns to help make customer selections.}
  \label{fig:sigma_workbook_interface}
\end{figure*}

\section{Introduction}
  Three major correlated forces act on enterprise data analysis today. First, more and more enterprise data are stored on the cloud. Second, the size and the diversity of data available to enterprises increase faster than ever as a growing number of data sources feed into enterprises' cloud data warehouses (CDWs). Third, the number of users in enterprises who would like direct access to cloud data for decision-making is also rapidly growing, boosting the demand for easy-to-use cloud data analysis systems such as Sigma Workbooks~\cite{DBLP:journals/pvldb/GaleSUFWD22}. Hence enabling easy and effective data discovery is increasingly more valuable. However, existing commercial solutions such as data catalogs are typically inaccessible from business intelligence tools, inadequate for surfacing relationships unknown to users, and inflexible in capturing data semantics. 

  Not surprisingly, recent years have seen a rising interest in data discovery, especially for Open Data~\cite{DBLP:journals/debu/MillerNZCPA18, DBLP:journals/pvldb/NargesianZPM18, DBLP:conf/sigmod/ZhuDNM19, DBLP:conf/icde/BogatuFP020}. This is a challenging problem due to the unstructured/semi-structured nature of Open Data and the lack of high-quality metadata. However, we observe that modern organizations also face data discovery challenges when dealing with structured data (i.e., databases in CDWs) for similar reasons: (i) in large CDWs, few people have a good understanding of data and their relationships; (ii) data is added in a much less modeled state as cloud computing power and storage become easily accessible; and (iii) business users who do not have comprehensive knowledge of underlying schema designs, are usually unaware of relationships between datasets (e.g., PK/FK relationship between two tables). There is a need for surfacing such relationships to help users connect related datasets within and across databases. 
  We also observe that CDW data tends to have more structure than Open Data, sometimes has high update rates, and usually has far more stringent requirements for completeness of discovery results.
  
  Consider a business user Joey who has access to a CDW that comprises hundreds of databases and thousands of tables (Figure~\ref{fig:sigma_workbook_interface}). In the interest of launching a sales campaign and finding potential customers, she starts with the \code{ACCOUNT} table in their \code{SALESFORCE} database, which contains member and billing data of their customer companies. To select ideal customers for the campaign, she wants to augment the \code{ACCOUNT} table with additional information such as sectors in which companies are engaged in their business activities. Now she has two options to find that complementary information: either she manually browses tables in the same database (even wanders through the entire data warehouse), or she can ask internal data teams for support. Either way, her workflow is significantly delayed. There is also a fair chance that she cannot go through all databases as it is a tedious and laborious process, and the data team does not have that particular piece of knowledge.

  Despite efforts in prototyping data discovery for real-world applications~\cite{DBLP:conf/icde/FernandezAKYMS18, DBLP:journals/pvldb/BharadwajGBG21, DBLP:conf/edbt/BogatuPDF22}, aspects of data discovery pertaining to enterprise settings are overlooked. For example, previous works assume a full pass of underlying data in the indexing/profiling phase. However, it is computationally and monetarily expensive to pull all data out of CDWs. In addition, existing solutions, including those for Open Data, focus on improving index lookup time for the scale of Open Data lakes or Web table corpora, which usually consist of a very large number (million scale) of small tables. In contrast, we observe that Sigma customers' CDWs have several thousand tables on average but can have billions of rows. Due to the relatively smaller number of tables but larger table sizes, index lookup time becomes less dominant compared to data loading time and profiling time, both of which can now bottleneck end-to-end query response time in the pipeline. So, the main challenge in the Sigma scenario shifts in part from searching over a massive number of small tables to optimizing the entire pipeline for a fair number of very large tables (e.g., reducing data loading time and using efficient profiling methods).
 
  In this work, we introduce a data discovery system \sysname and share our progress. \sysname prototypes semantic join discovery in Sigma, an enterprise platform for cloud data analytics. Specifically, we make the following contributions:
  \begin{itemize}
    \item We present \sysname, a semantic join discovery system, which demonstrates a real-world usage scenario of data discovery within an enterprise product.
    \item We identify unique challenges in Sigma and propose an embedding-based solution to capture semantic relationships between tables, especially those across databases. 
    \item We evaluate \sysname on three cross-domain table corpora, including one collected from a CDW, and show our embedding-based approach is effective and robust to sampling. We also discuss further optimization opportunities for deployment and larger-scale discovery.
  \end{itemize}

\section{Problem Definition}\label{sec:prob_def}
  In this section, we first briefly introduce Sigma and describe the data discovery needs of Sigma users. Then we formally define our problem.

\subsection{Data Discovery Need in Sigma}
  Sigma is a web-based SaaS platform for cloud data analytics and business intelligence. It features Sigma Workbooks, which provides a spreadsheet-like interface for business users to perform interactive visual analysis of data in CDWs. 
  
  Among many features familiar to business users who are more skilled in spreadsheet functions than SQL, \code{Lookup} enables users to write a spreadsheet formula and add a column from another data source (i.e., another table) to enrich the dataset at hand. Under the hood, Workbooks translates spreadsheet formulas into SQL queries required to fetch data from cloud databases and performs cardinality-preserving joins on related tables.
  
   A major limitation of \code{Lookup} is that users have to specify the join path in formulas. In other words, it requires users to have prior knowledge about relationships between datasets. However, users often lack such knowledge and have to rely on their data teams for help or manually browse tables in databases to find a clue. This limitation has become a bottleneck in fully utilizing the \code{Lookup} functionality, significantly delaying their data exploration and analysis workflows. We believe this sort of data discovery will be a challenge for all cloud analytics products and hence has to be addressed. 

\subsection{Semantic Join Discovery}
  \begin{ps}[Top-$k$ Semantic Join Discovery]
    Given a corpus of tables $\mathcal{S}$, a query column $\mathbf{c_{q}}$ from a table $\mathbf{Q}$, and a constant $k$, find up to $k$ candidate columns from $\mathcal{S}$ such that they are most likely to be joinable with $\mathbf{c_{q}}$.
  \end{ps}

  Traditionally, joinable tables have been discovered through syntactic analysis, which can miss many join opportunities, particularly when data are independently sourced. High dimensional embeddings have recently been shown to capture semantics effectively in many domains~\cite{DBLP:journals/pvldb/DengSL0020, DBLP:journals/pvldb/ZhangSLHDT20}. We expect that the semantic similarity of two columns can be quantified by some similarity measure between their embeddings, which we expect can serve as a proxy for their join-ability as well.

  \begin{definition}[Semantic Column Join-ability]\label{def: sem_col_join}
    Given two columns $A$ and $B$, we define semantic column join-ability $\mathcal{J}(A, B)$ as $$\mathcal{J}(A, B) = \mathcal{M}(\mathcal{T}(A), \mathcal{T}(B))$$ where $\mathcal{T}(\cdot)$ is an embedding function that projects a given column into a vector space and $\mathcal{M}(\cdot, \cdot)$ is a similarity measure between two embeddings in the vector space.
  \end{definition}
  
  Based on this metric, we can attempt to solve the semantic join discovery problem by returning candidate tables in descending order of semantic column join-ability $\mathcal{J}$ relative to the query column $\mathbf{c_{q}}$. How to do this effectively is the central challenge addressed in this paper.
    
\section{\plainsysname Overview}
  \begin{figure}[h!t]
    \centering
    \includegraphics[width=0.9\linewidth]{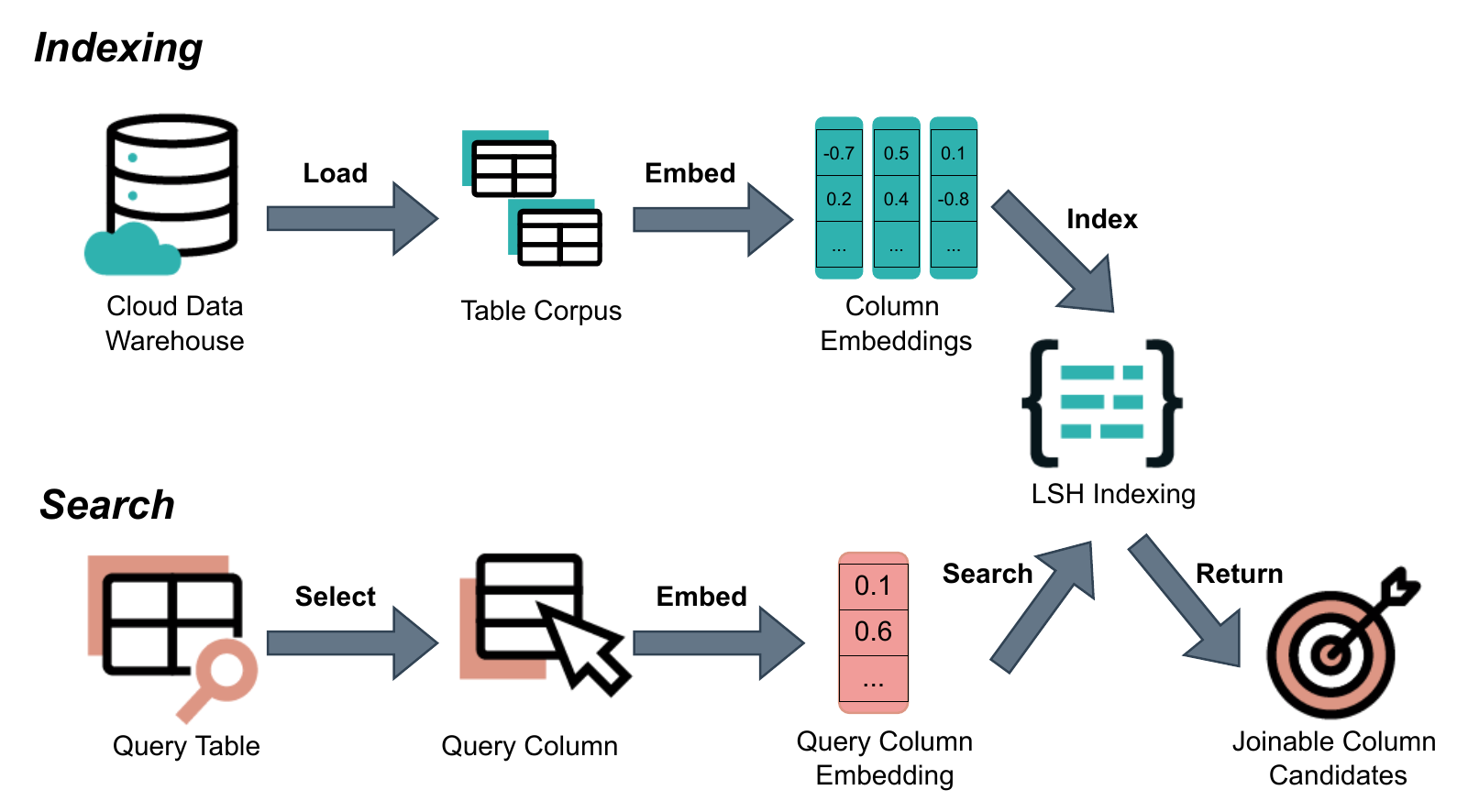}
    \caption{\sysname pipeline.}
    \label{fig:warpgate_pipeline}
  \end{figure}

  We discuss key challenges to implementing an embedding-based approach to the semantic join discovery problem, along with our solutions to these challenges in Section~\ref{subsec:warpgate_prototype}. Then we describe our system prototype \sysname incorporating these ideas in Section~\ref{subsec:warpgate_interface}.

\subsection{Embedding-Based Solution}\label{subsec:warpgate_prototype}
  Figure~\ref{fig:warpgate_pipeline} gives an overview of the \sysname architecture, which consists of two main pipelines: indexing and search, which we consider in turn below. But first, we discuss the embedding step common to both pipelines.

\subsubsection{Column Embeddings}
  We consider embeddings at the column level an effective encapsulation of semantics for join discovery. Intuitively, embeddings can capture semantic relationships between columns across databases where value formatting may vary widely despite semantic proximity.

  However, the effectiveness of embeddings depends on the embedding model. We consider several factors in choosing an embedding model:
  \begin{enumerate}[label=(\roman*)]
    \item Whether the model is (pre-)trained over tabular data;
    \item The size of the training corpus;
    \item The efficiency of model inference.
  \end{enumerate}

  We observe two primary paradigms in applying embeddings for data management tasks. Early work ~\cite{DBLP:conf/sigmod/BordawekarS17} leverages off-the-shelf embedding models from the natural language processing (NLP) field. Recent work~\cite{DBLP:conf/sigmod/0002TGL21, DBLP:journals/pvldb/DengSL0020} adapts NLP model architectures and training algorithms for tabular data. As demonstrated in their work, embeddings from models trained on tabular data manifest better performance in downstream tasks. For instance, using the same training algorithm, ~\cite{DBLP:conf/sigmod/0002TGL21} shows that models trained on sequences extracted from tables outperform the model trained on an unstructured text corpus. The performance uplift is attributed to the serialization of tabular data, which considers the table structure in the embedding process. With this observation, we prefer embedding models designed for tabular data.

  Empirically, more training data usually lead to better model performance and generalizability, given enough model capacity. In the database community, Web table corpora~\cite{DBLP:conf/ssdbm/EberiusTBL15, DBLP:conf/www/LehmbergRMB16} extracted from the Common Crawl contribute the largest publicly accessible relational table datasets. Efforts exploring representation learning for relational tables~\cite{DBLP:conf/sigmod/0002TGL21, DBLP:journals/pvldb/DengSL0020} have trained embeddings on various Web table corpus, which also exhibits the impact of Web tables beyond Web applications. Thus, we favor embedding models derived from those large table corpora.

  The embedding process for tabular data generally involves table serialization, input encoding (e.g., tokenization and transformation), and model inference. It is worth noting that although more complex models (in terms of the number of parameters) give embeddings that tend to yield better results in the downstream tasks, they also carry a larger inference cost (i.e., longer inference time, which directly affects query response time) compared to simple models. For interactive applications such as join discovery and recommendation in Sigma, we need to choose models that balance the embedding quality and model inference time.

\subsubsection{Indexing}
  We employ locality-sensitive hashing (LSH)~\cite{DBLP:conf/vldb/GionisIM99, DBLP:conf/stoc/Charikar02} to turn the high-dimensional embedding similarity search problem into nearest neighbor search in a low-dimensional space. The basic idea of LSH is to maximize hash collisions for similar inputs. With respect to a similarity measure, LSH applies a specific family of hash functions that assign two similar inputs to the same ``buckets'' with high probability that is equal to the similarity score between them.

  In our case, we consider the semantic join-ability of two columns to be the cosine similarity of their embeddings. To approximate the cosine similarity of two vectors, we rely on SimHash (also known as random projection)~\cite{DBLP:conf/stoc/Charikar02} that essentially uses hyperplanes in the vector space as hash functions. We hash all column embeddings from the table corpus into a SimHash LSH index. When a query comes in, the LSH index hashes the query embedding and only searches within the sub-universe of embedding vectors (much smaller than the entire universe) that share the same hash as the query, hence reducing the search time.

\subsubsection{Sampling}
  Existing data discovery systems~\cite{DBLP:conf/icde/FernandezAKYMS18, DBLP:conf/icde/BogatuFP020, DBLP:conf/edbt/BogatuPDF22} run a one-pass data scanning step to read all datasets once and retrieve profiling information (e.g., MinHash for each column). Reading full tables of millions of rows or even billions of rows from CDWs is expensive in terms of both computation time and monetary cost because CDW vendors that adopt a usage-based pricing model (i.e., pay-as-you-go) charge per GB of data scanned. It is then intuitive to use sampling to reduce the cost. But one side effect of sampling is potential disruptions to column profiles. For instance, profiling methods like MinHash are shown to be sensitive to the sample size~\cite{DBLP:journals/pvldb/ChenNC14}. It also remains unclear what a good sample size is for embeddings to preserve the similarity for data discovery.

  \begin{figure*}[h!t]
    \centering
    \includegraphics[width=0.95\linewidth]{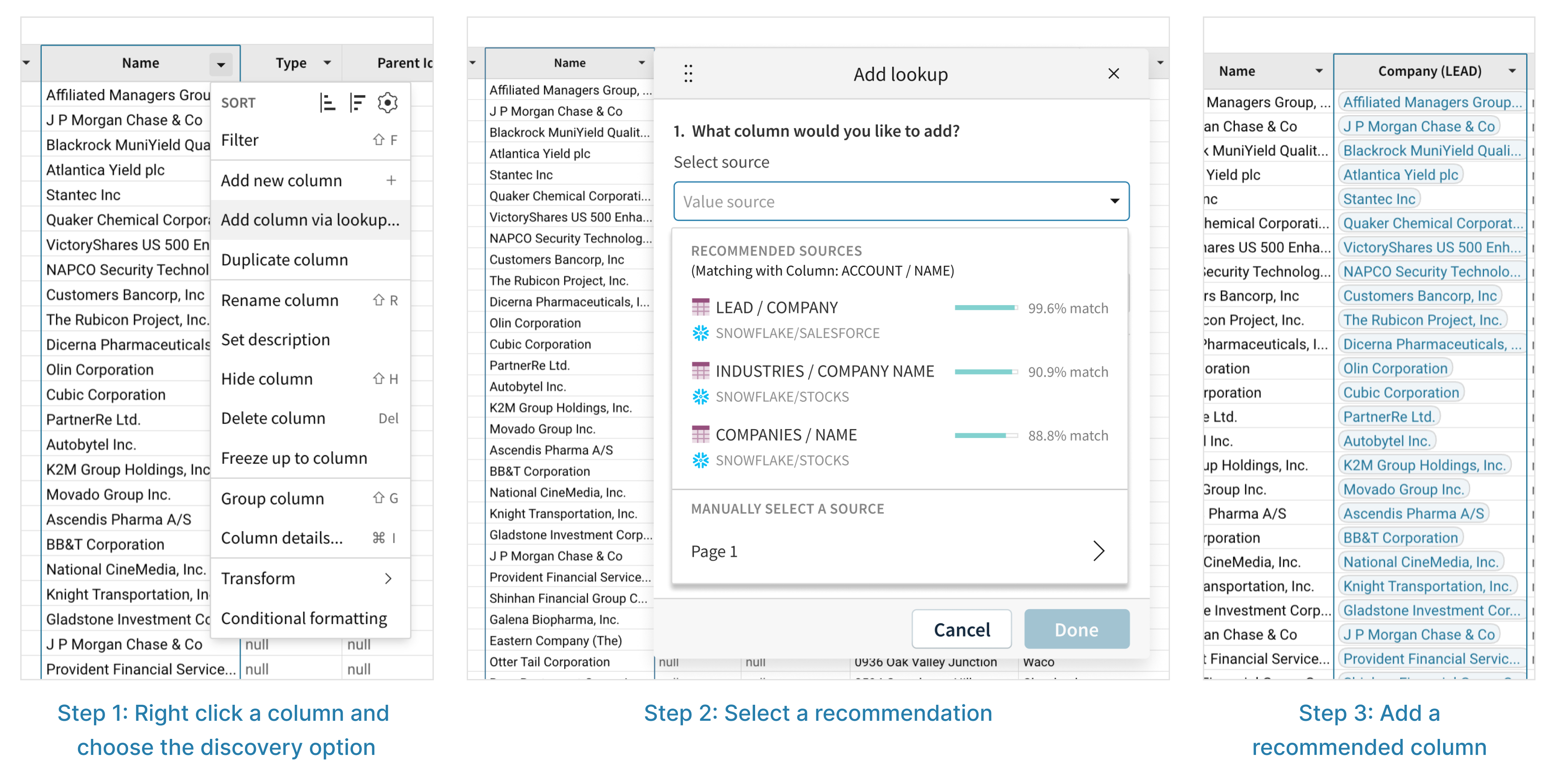}
    \caption{\sysname is integrated into a functionality called \code{Add column via lookup} in Sigma Workbooks.}
    \label{fig:interface_functionality}
  \end{figure*}

\subsection{\plainsysname Interface}\label{subsec:warpgate_interface}
  We have implemented the ideas described above in a prototype system we call \sysname. As shown in Figure~\ref{fig:interface_functionality}, we integrate \sysname into Sigma Workbooks as a functionality called \code{Add column via lookup}.
  
  Continuing the example of a business user Joey who wants to identify customers for a sales campaign, she opens up the \code{Account} table and sees a full list of companies (the \code{Name} column highlighted in Step 1 of Figure~\ref{fig:interface_functionality}). To help make the decision, she wants to augment the table with additional information so that she can know which companies fall into the scope of the sale campaign. In the current interface, she right-clicks on the \code{Name} column and chooses the \code{Add column via lookup} option in the menu. A window shows up (Step 2 in Figure~\ref{fig:interface_functionality}) and displays top-$k$ join path recommendations. Each recommendation gives the candidate join column, the table, and the database from which the candidate column comes, as well as a similarity score relative to the \code{Name} column (i.e., the query column). Although recommendations are ranked in descending order of the similarity score, Joey can pick one that fits her interest most. Once she chooses a candidate, the window will display a list of all the columns from the candidate table. Joey can browse and select the most relevant columns. For now, the interface will add user-selected columns to the \code{Account} table right at the side of the \code{Name} column for information complement and also load the candidate table into Sigma Workbooks for further inspection. In Step 3, we simply add and show the first recommended column, which is a foreign key in the same database.
  
  The current interface design is based on the consideration that it is less desirable to simply join two wide tables with many columns and overwhelm users with a display of an even wider table that does not really help users with their information needs. In this case, Joey can first add the most relevant columns at first glance based on her domain knowledge while having the flexibility to see the entire candidate table for more careful examination.

\section{Experiments}
  We evaluate \sysname on effectiveness and efficiency in comparison with two existing data discovery systems that support enterprise use cases. We obtain promising results of our embedding-based approach and show that it is sample-efficient and scales well to tables of millions of rows.

\subsection{Datasets}
  We use three repositories for evaluation. Table~\ref{tab:datasets_stats} gives a summary of dataset characteristics.

\begin{table}[h!]
\caption{Basic statistics of evaluation datasets. XS is shorthand for testbedXS in the NextiaJD repository, similarly for S, M, and L.}
\label{tab:datasets_stats}
\resizebox{0.95\columnwidth}{!}{%
\begin{tabular}{cccccc}
\hline
\multicolumn{1}{l}{}     & \multicolumn{1}{l}{\textbf{\# Tables}} & \multicolumn{1}{l}{\textbf{\# Columns}} & \multicolumn{1}{l}{\textbf{Avg. \# Rows}} & \multicolumn{1}{l}{\textbf{\# Queries}} & \multicolumn{1}{l}{\textbf{Avg. \# Answers}} \\ \hline
\textit{\textbf{XS}}     & 28                                     & 257                                     & 1,938                                     & 35                                      & 2.8                                          \\ \hline
\textit{\textbf{S}}      & 46                                     & 2,553                                   & 209,646                                   & 177                                     & 3.6                                          \\ \hline
\textit{\textbf{M}}      & 46                                     & 1,067                                   & 3,175,904                                 & 188                                     & 4.4                                          \\ \hline
\textit{\textbf{L}}      & 19                                     & 541                                     & 12,288,165                                & 92                                      & 3.6                                          \\ \hline
\textit{\textbf{Spider}} & 70                                     & 429                                     & 7632                                      & 60                                      & 1.1                                          \\ \hline
\textit{\textbf{Sigma}}  & 98                                     & 1,343                                   & 2,243,932                                 & TBD                                     & N/A                                          \\ \hline
\end{tabular}%
}
\end{table}

  \subhead{\nextiaJD.} Flores et al.~\cite{DBLP:conf/edbt/0002N021} composes four testbeds of datasets from open repositories such as Kaggle and OpenML. Datasets are divided into testbeds according to their file size. For instance, \nextiaJD-XS contains datasets of size smaller than 1 MB while \nextiaJD-L consists of datasets of size larger than 1 GB. They also label the join quality of pairs of attributes based on a measure that considers both containment and cardinality proportion with empirically determined thresholds. In experiments, we use attribute pairs with quality labeled as Good and High by ~\cite{DBLP:conf/edbt/0002N021}.
  
  \subhead{Spider.} Released as a large-scale semantic parsing and text-to-SQL dataset, Spider~\cite{DBLP:conf/emnlp/YuZYYWLMLYRZR18} includes 5,693 SQL queries on 200 databases across domains. We parse schema SQL files and retrieve join paths between primary keys and foreign keys as ground truth. Spider has both the training set and the development set, and we currently use join paths from the development set for evaluation. 
  
  \subhead{Sigma.} The Sigma Sample Database is a collection of schema and tables that are accessible to all Sigma Computing accounts for exploration.  It is stored in a Snowflake account managed by Sigma Computing and changes over time.  The datasets in this database vary in origin; some are real, publicly accessible data, some are real data that has been obfuscated, and some are completely auto-generated data. The domain of the data ranges includes retail (transactions, products and stores), financial (daily attributes of many securities), demographic (census, restaurants and bikes) and usage (cloud application usage and metering, server logs).

\subsection{Baselines \& Metrics}
  For comparison, we consider as baselines two system prototypes  Aurum~\cite{DBLP:conf/icde/FernandezAKYMS18} and {D$^{3}$L}~\cite{DBLP:conf/icde/BogatuFP020} that report on real-world data discovery (see Section~\ref{sec:related_work} for more information). Note that Aurum stores detected relationships in a graph data structure and does not support top-$k$ search.

  When ground truth is available, we report top-$k$ precision and recall with varying $k$. At each value of $k$, we average the precision and recall numbers over all the queries. We also limit $k$ to a small range since the average number of answers to a query is small in evaluation datasets, and we don't want to overwhelm users with too many recommendations.
  
  Besides effectiveness, we consider index lookup time and end-to-end query response time for efficiency measurement. Index lookup time refers to the time for the index data structure to return top-k candidates, whereas end-to-end query response time is the total amount of time for a discovery system to respond to a query, which includes data loading and embedding inference time in our case. We report both metrics in seconds averaged over all queries in a dataset. It is crucial in our application scenario that the system responds to queries at interactive speed.

\subsection{Experiment Results}
   Based on considerations in Section~\ref{subsec:warpgate_prototype}, we choose to use Web Table Embeddings~\cite{DBLP:conf/sigmod/0002TGL21} as our underlying embedding model. We set the similarity threshold of the SimHash LSH index to 0.7 and run baselines with their default setting. All experiments are conducted on an Amazon EC2 p3.8xlarge instance.

\subsubsection{Join Discovery on nextiaJD Testbeds}
  \begin{figure*}[!ht]
    \centering
    \minipage{0.33\linewidth}
      \includegraphics[width=\linewidth]{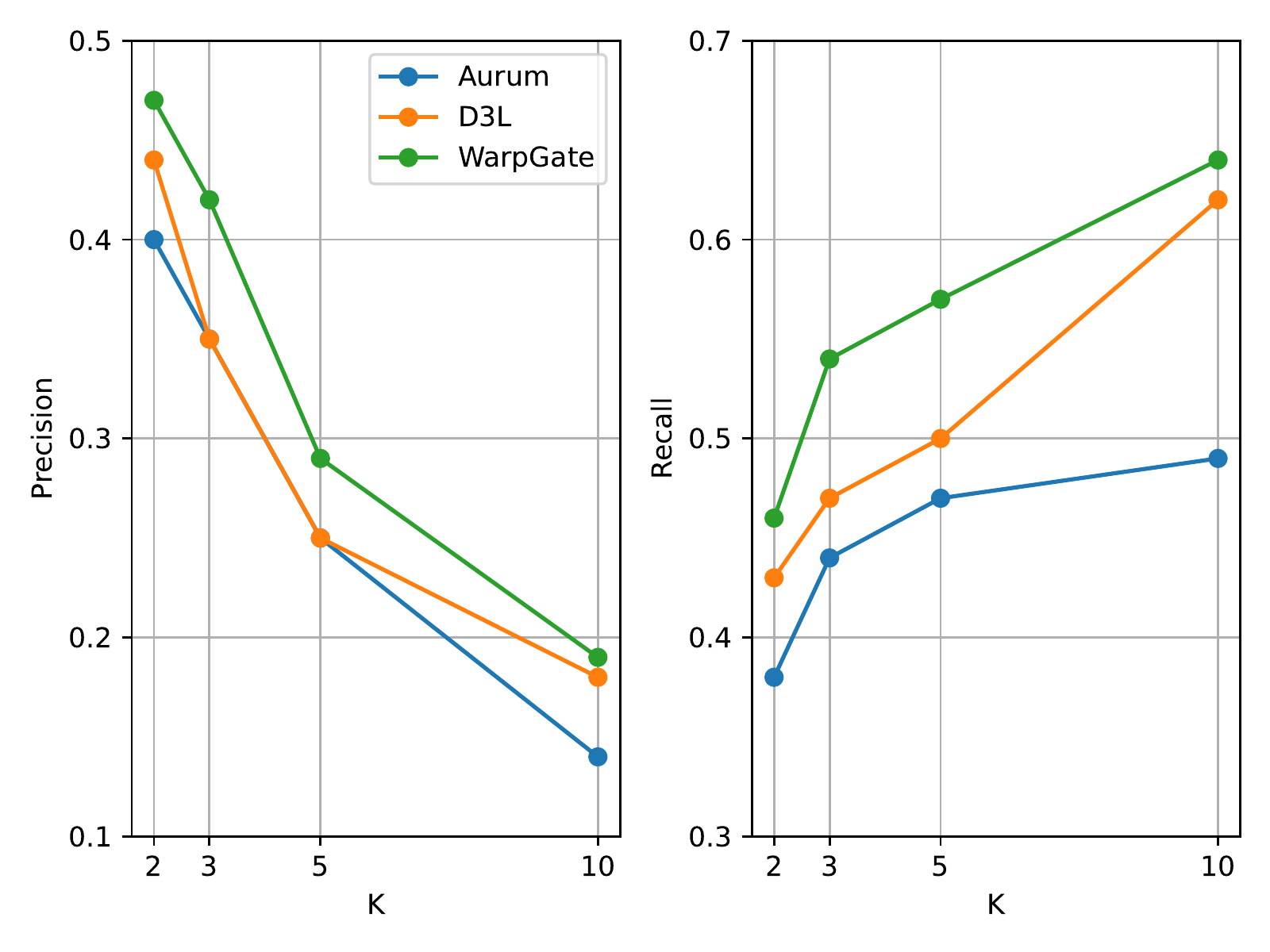}
      \vspace{-0.8cm}
      \caption*{(a) testbedS}
    \endminipage\hfill
    \minipage{0.33\linewidth}
      \includegraphics[width=\textwidth]{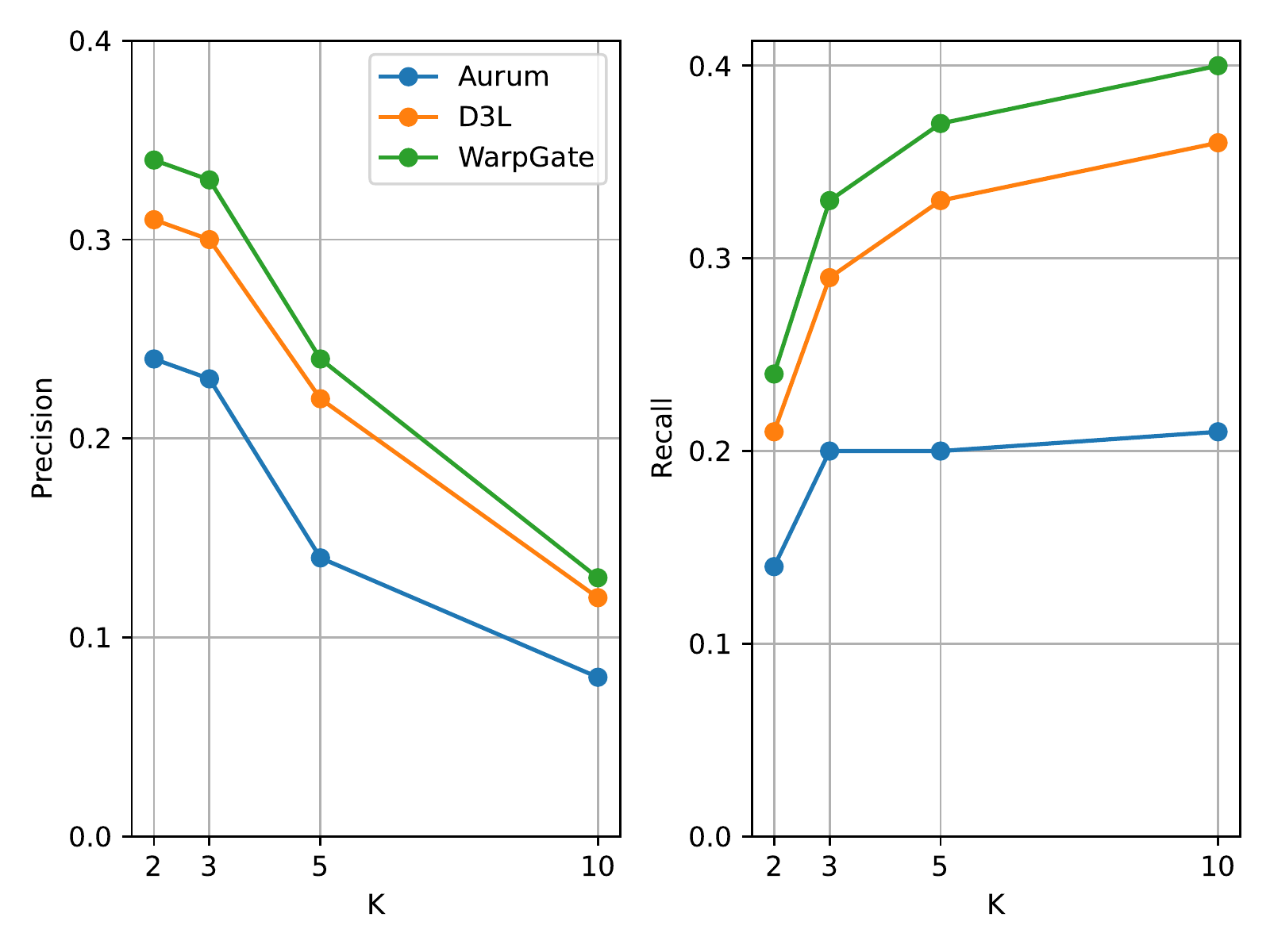}
      \vspace{-0.8cm}
      \caption*{(b) testbedM}
    \endminipage\hfill
    \minipage{0.33\linewidth}
      \includegraphics[width=\textwidth]{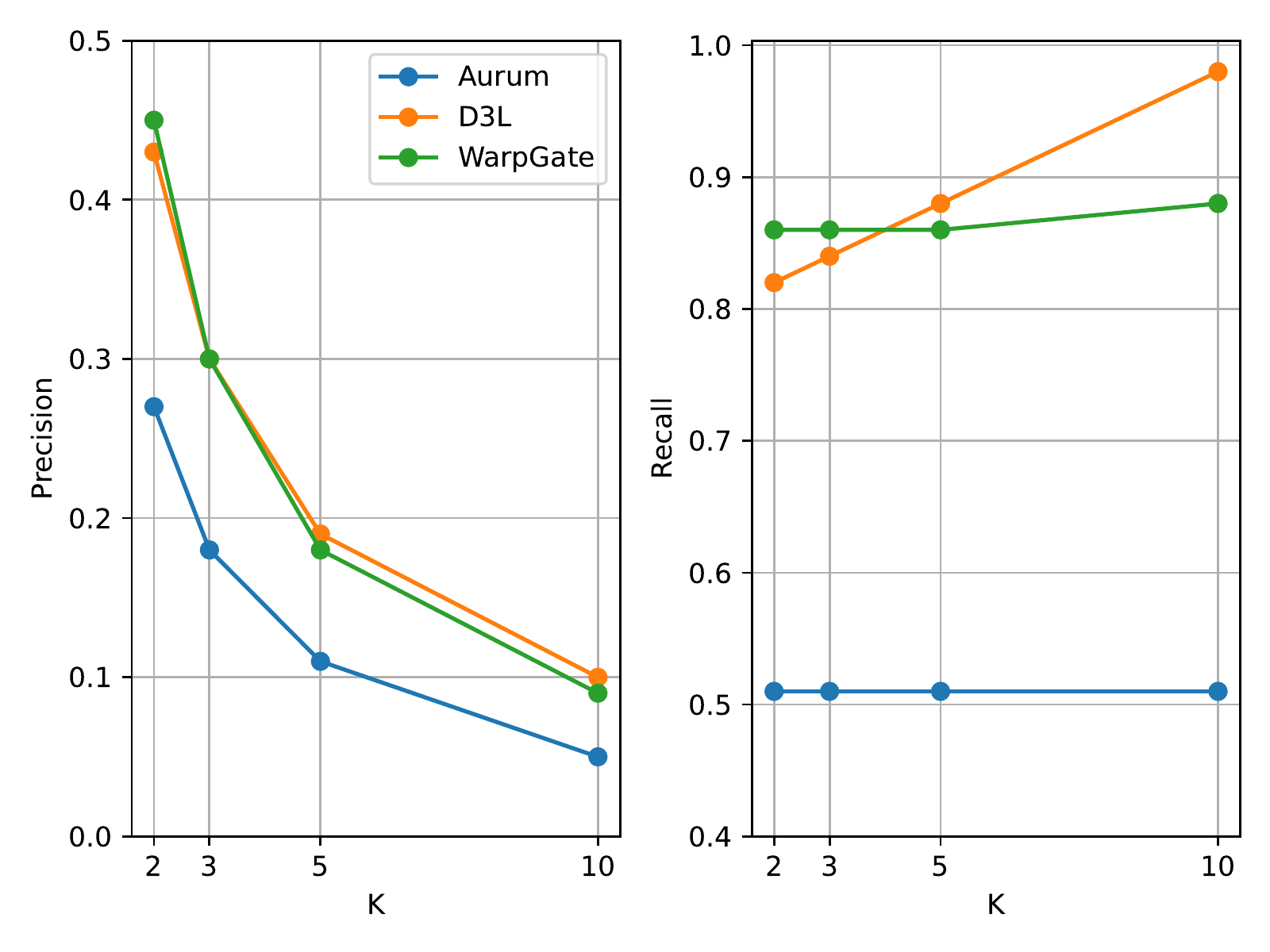}
      \vspace{-0.8cm}
      \caption*{(c) Spider}
    \endminipage\hfill
    \caption{Top-$k$ precision and recall.\label{fig:quality_nextiaJD_spider}}
    \vspace{-1em}
  \end{figure*}
  
  We present discovery results on \nextiaJD testbedS and testbedM in Figure~\ref{fig:quality_nextiaJD_spider}(a) and ~\ref{fig:quality_nextiaJD_spider}(b) respectively (results on four testbeds have similar trends while these two testbeds contain more tables and queries than the other two). As shown, \sysname consistently obtains higher precision and recall compared with two baselines as $k$ increases.

  \begin{small}
  \begin{table}[h!]
    \centering
    \caption{End-to-end query response time in seconds/query (for $k=10$) along with index lookup time reported in paratheses for \sysname.}
    \label{tab:query_response_time}
    \resizebox{0.7\columnwidth}{!}{%
    \begin{tabular}{cccc}
    \hline
                               & \textbf{Aurum} & \textbf{D$^{3}$L} & \sysname \\ \hline
    \textit{\textbf{testbedS}} & 0.18           & 4.77         & 3.12 (1.04)             \\ \hline
    \textit{\textbf{testbedM}} & 0.03           & 57.69        & 38.73 (8.39)            \\ \hline
    \end{tabular}%
    }
  \end{table}
  \end{small}

  Table~\ref{tab:query_response_time} shows end-to-end query response time. Aurum is significantly faster because it indexes detected relationships in another graph data structure for result retrieval. In contrast, D$^{3}L$ is the slowest since it uses an ensemble approach and aggregating multiple types of signals takes longer. As we can also see here, given the equal number of tables in two testbeds, when the average number of rows increases by an order of magnitude, query response time increases linearly and becomes unacceptable for interactive purposes. Moreover, index lookup time accounts for less than 25\% of query response time on testbedS and less than 13\% on testbedM. This suggests that there are other bottlenecks in the pipeline and improving only index lookup time (as many earlier works focus on) is not enough to reduce overall query response time. Upon inspection, we found loading large datasets from disk to memory and embedding inference take most of the time besides index lookup. We show in Section~\ref{subsec:sample_efficiency} that sampling is an effective mitigation and our embedding approach is robust to sample size.

\subsubsection{PK/FK Detection on Spider}
  Figure~\ref{fig:quality_nextiaJD_spider}(c) shows the top-$k$ precision and recall of three systems on the Spider dataset. We use this dataset mainly to demonstrate that for PK/FK detection within each independent database, the embedding measure alone (\sysname) can compare favorably against the ensemble approach (D$^{3}$L) and outperform the syntactic-only approach (Aurum) by a large margin. As to D$^{3}$L, the jump of recall from $k=5$ to $k=10$ is attributed to the column name similarity measure in D$^{3}$L, and we indeed observe in Spider that many PK/FKs share syntactically similar column names. Since Spider has a small number of tables that contain thousands of rows on average, the search time is fast for all three systems, which is no more than 2 seconds for running all the queries.

\subsubsection{Ad-Hoc Discovery in Sigma}
  Since Sigma Sample Database is a corpus without ground truth, we ask four colleagues to pick columns of interest as queries. Here we describe one of the most interesting discoveries. In future work, we plan to conduct a user study to evaluate the value of \sysname in a principled manner.

  Continuing our running example of the business user Joey, who is interested in looking for additional information to help make the decision of customer selection. Figure~\ref{fig:interface_functionality} displays the top-3 recommendations from \sysname. The first candidate is the \code{Company} column of the \code{LEAD} table from the same database as the query column, which contains information on contact points in each company (e.g., name, title, and address of each contact point). Upon browsing columns in the \code{LEAD} table, Joey does not find auxiliary information for her need. She moves to inspect the second candidate, which is the \code{Company Name} column of the \code{INDUSTRIES} table from the \code{STOCKS} database. In the same table, the \code{Industry Group} column has sector information of US public companies, which can (partially) enrich the query table. Joey then acts on her customer selection by filtering companies with their sectors. Even more interestingly, she can add to the query table the \code{TICKER} column from the \code{INDUSTRIES} table, which can be further used as the join key to add columns of stock prices from the same \code{STOCKS} database. This way, Joey can track down high-performing companies in targeted sectors as her sales campaign customers.

\subsection{Sample Efficiency}\label{subsec:sample_efficiency}
  We run \sysname over \nextiaJD-S and \nextiaJD-M with sample size 10, 100, and 1000. For all sample sizes, the embedding approach remains as effective as using full values with $\pm 1\% / 2\%$ variation at different $k$ values. In the meanwhile, index lookup time is reduced by up to two orders of magnitude (e.g., from about 1 second to about 10 milliseconds on \nextiaJD-S), and end-to-end query response time also comes down to interactive speed (under 35 milliseconds per query on \nextiaJD-S and under 65 milliseconds per query on \nextiaJD-M). We discuss further optimization opportunities in Section~\ref{sec:discussion}.

  To see if sample efficiency is unique to Web Table Embeddings, we also evaluate BERT~\cite{DBLP:conf/naacl/DevlinCLT19} as our underlying embedding model. We observe that the effectiveness of BERT Embeddings is robust to sample size as well. Although BERT is a significantly more complex model, we do not see much improvement in effectiveness (mostly on par with Web Table Embeddings). Nevertheless, index lookup time and query response time are 10x slower without sampling due to more expensive embedding inference costs.
\section{Discussion}\label{sec:discussion}
\subsection{Sigma Customer Data Scale} The scale of enterprise CDWs is a challenge for any deployment of join discovery. The median Sigma customer warehouse has 450 tables for analysis, but the average is over 12,700 tables with an average of 25.7 columns per table. The size of enterprise tables is also a challenge. The median table in these warehouses has just 7,700 rows, but the average is 1.7 billion rows. Actively sampling this large number of tables incurs usage costs in CDWs, so a passive sampling of user queries is beneficial. Likewise, samples can be shared across ML applications in Sigma to amortize their cost.

\vspace{-1em}
\subsection{Optimization Opportunities}
  \subsubsection{Contextual Embeddings.} Existing approaches generally follow a two-step process: they first build profiles for each column independently and then determine join-ability of two columns based on some notion of similarity between their profiles (e.g., containment or Jaccard similarity). Typically, the context of a column is not taken into account when building those profiles. However, context (e.g., other columns in the same table, user activities, query logs) can potentially provide auxiliary information that is critical to find semantically related candidates. We plan to explore the option of incorporating context information into the underlying embedding model of \sysname.

  \vspace{-1em}
  \subsubsection{Data Management for Data Discovery.} CSV is a common file format for tabular data in Open Data Lakes and table repositories on the Web. Despite the flexibility and human readability of CSV files, the flat-file format is not ideal for storage and query purposes. We note that (1) loading giant CSV files and running embedding inference can incur heavy memory usage and crash programs on small machines; (2) operations in join discovery like embedding inference are column-oriented; (3) what is unique in our scenario is that our data are pulled from CDWs, thus more structured and much better data quality than those in the wild. (1) and (2) make us think it is more memory and storage efficient to take advantage of (in-memory) column stores for data pulled out of CDWs, and (3) can make the process easier. This change will also have significant implications for deployment as we need to provision CPU and memory usage in Sigma's Kubernetes cluster. 

  \vspace{-1em}
  \subsubsection{Efficient Search.} To further improve query response time for Sigma customers with a massive number of tables in their CDWs, we are considering two directions: (1) a block-and-verify strategy in ~\cite{DBLP:conf/icde/DongT0O21}, which employs pivot-based filtering to prune dissimilar vectors; (2) fine-tune off-the-shelf embedding models in a self-supervised way that pushes embeddings of joinable columns to have higher cosine similarity so that an index data structure like SimHash can be better utilized~\cite{cong2022pylon}.
\section{Related Work}\label{sec:related_work}
 \sysname builds on earlier work on data discovery for enterprise data systems and Open Data, which we summarize below. We distinguish \sysname from this earlier work mainly in two aspects: (i) we recognize and showcase the value of data discovery for structured data in enterprise CDWs; (ii) we do not assume a full pass of data in the indexing/profiling phase and show that the embedding-based approach is effective and robust to data samples.

\subhead{Data Discovery Systems for Enterprise.} Aurum~\cite{DBLP:conf/icde/FernandezAKYMS18}  models syntactic relationships between datasets in a graph data structure to support data discovery. In a two-step process of profiling data and computing the relationship between data signatures, Aurum builds up a graph with nodes representing column profiles and weighted edges indicating syntactic relationships between two nodes (e.g., content similarity). With an in-memory index of the graph, Aurum can efficiently support various discovery needs, such as how similar two columns are, which can be used to find joinable datasets. 
 SemProp~\cite{DBLP:conf/icde/FernandezMQEIMO18} extends Aurum by using word embeddings to surface semantically related objects in the graph.

  Voyager~\cite{DBLP:conf/edbt/BogatuPDF22} built on top of D$^{3}$L~\cite{DBLP:conf/icde/BogatuFP020} formalizes and supports three tasks routinely performed by data scientists using their platform. It profiles data with a full scan and indexes columns with MinHash. Voyager also drops the embedding measure in D$^{3}$L and relies only on syntactic measures.

\subhead{Join Discovery over Open Data.} JOSIE~\cite{DBLP:conf/sigmod/ZhuDNM19} considers the problem of finding joinable tables in modern data lakes. It formulates the problem as a top-k overlap set similarity search problem where columns are treated as sets and matching values as intersections between sets. By using a cost model for estimating the reading costs of data structures, JOSIE contributes an exact overlap set similarity search algorithm that can adapt to different data distributions. 

  Juneu~\cite{DBLP:conf/sigmod/ZhangI20} aims to help users find semantically related datasets in data lakes to augment data analytics tasks, including finding joinable tables in interactive data science environments (e.g., Jupyter Notebook). It combines multiple table-relatedness measures such as column and row overlap, provenance, and user context and intent for related-table search. In a similar vein, D$^{3}$L~\cite{DBLP:conf/icde/BogatuFP020} uses five types of evidence to determine column proximity: (i) column name similarity; (ii) column extent overlap; (iii) word-embedding similarity; (iv) format representation similarity; and (v) domain distribution similarity for numerical attributes. This ensemble approach is shown to be more effective and efficient than Aurum on table union search, which is a problem that has significant overlap with join path discovery.

  In contrast to the search paradigm, Nargesian et al.~\cite{DBLP:conf/sigmod/NargesianPZBM20} propose building a navigation graph over data lakes. By leveraging a probabilistic model, they show that navigation can help users find relevant tables that cannot be found by keyword search.

\section{Conclusion}
  We present \sysname, a system prototype for semantic join discovery in Sigma Workbooks. In future work, we expect to share our experience and lessons of deploying \sysname in the production environment and conduct a user study to further evaluate its value for end users.

\begin{acks}
  We thank Oscar Bashaw, Nipurn Doshior, Qing Feng, Jade Fleishhacker, and Siva Obulam for their help in \sysname's development. This project is supported in part by NSF grant 1946932.
\end{acks}
\bibliographystyle{ACM-Reference-Format}
\bibliography{sections/reference.bib}

\end{document}